\documentclass[prl,showpacs,reprint]{revtex4-1}
\pdfoutput=1

\usepackage[utf8]{inputenc}
\usepackage{hyperref}
\usepackage{amsmath}
\usepackage{amsfonts}
\usepackage{amssymb}
\usepackage{graphicx}

\DeclareMathOperator\tr{tr}

\newcommand{\Sdet}{\sigma}
\newcommand{\tphi}{\psi}
\newcommand{\Ns}{N_s}
\newcommand{\Y}{O}
\newcommand{\D}[2]{D_{#1,#2}}

\begin{document}

\title{Evading the sign problem in random matrix simulations}

\author{Jacques Bloch}
\affiliation{Institute for Theoretical Physics, University of Regensburg, 93040  Regensburg, Germany}

\email{jacques.bloch@physik.uni-regensburg.de}

\date{March 17, 2011}


\begin{abstract}
We show how the sign problem occurring in dynamical simulations of random matrices at nonzero chemical potential can be avoided by judiciously combining matrices into subsets. For each subset the sum of fermionic determinants is real and positive such that importance sampling can be used in Monte Carlo simulations. The number of matrices per subset is proportional to the matrix dimension. We measure the chiral condensate and observe that the statistical error is independent of the chemical potential and grows linearly with the matrix dimension, which contrasts strongly with its exponential growth in reweighting methods.
\end{abstract}


\maketitle

Dynamical simulations of QCD at nonzero chemical potential are hampered by the sign problem (see ref.~\cite{deForcrand:2010ys} for a review). The problem occurs when the probabilistic weight used in Monte Carlo methods becomes complex, thus precluding the use of standard importance sampling techniques.  Methods to study QCD at small chemical potential, where the sign problem is mild, include reweighting methods, Taylor expansions and analytic continuation from imaginary chemical potential \cite{deForcrand:2010ys}. QCD at zero and nonzero density can also be investigated using random matrix theory because of the universality properties of the Dirac operator to leading order in the epsilon regime.
Based on this equivalence, the severity of the sign problem has been studied using the average phase of the fermion determinant \cite{Splittorff:2006fu,Splittorff:2007ck,Bloch:2008cf,Lombardo:2009aw,Bloch:2011jk}. As dynamical simulations of random matrices at nonzero chemical potential also suffer from the sign problem they can be used as a playground for its study. In this letter we present a subset method which avoids the sign problem in this case.

In chiral random matrix theory at nonzero chemical potential $\mu$ the Dirac operator for a fermion of mass $m$ can be represented by \cite{Osborn:2004rf}
\begin{align} \D{\mu}{m}(\phi) =
  \begin{pmatrix}
    m & i\phi_1 + \mu\phi_2 \\
    i \phi_1^\dagger + \mu\phi_2^\dagger & m
  \end{pmatrix} ,
  \label{Ddef}
\end{align}
where the configurations $\phi=(\phi_1,\phi_2)$ are pairs of complex random matrices $\phi_1$ and $\phi_2$ 
of dimension $(N+\nu) \times N$ with $\nu$ the number of zero modes of $D$ (for $m=0$).
The partition function of the chiral random matrix model with $N_f$ dynamical quarks with equal mass $m$ is
\begin{align}
  Z = \int d\phi_1 d\phi_2 \, w(\phi_1) w(\phi_2) \,
  {\det}^{N_f} \D{\mu}{m}(\phi) ,
  \label{partfun} 
\end{align}
with Gaussian weights 
\begin{align} 
  w(\phi_i) = (N/\pi)^{N(N+\nu)} \exp(-N \tr \phi_i^\dagger \phi_i)
\label{rmtdis}
\end{align} 
and integration over the real and imaginary parts of all matrix components. 
In the presence of a chemical potential the fermion determinant becomes complex, the sign problem arises and the work needed to make reliable measurements on the statistical ensemble grows exponentially with the volume. 
In reweighting methods this exponential increase comes from the need to compute exponentially small reweighting factors from a statistical sampling of largely canceling contributions \cite{deForcrand:2010ys}.  

Herein we propose a method to avoid the sign problem in dynamical simulations of random matrices. The idea is to rewrite the partition function \eqref{partfun} as an integral over subsets $\Omega=\{\phi^i: i=1,\ldots,\Ns\}$, each containing $\Ns$ configurations $\phi^i=(\phi^i_1,\phi^i_2)$, such that
\begin{align}
  Z = \int d\Omega \, W(\Omega) {\Sdet_{\mu,m}}(\Omega) , 
  \label{Zsubset}
\end{align}
where by construction the Gaussian weights $w(\phi^i_1) w(\phi^i_2)$ will be invariant for all $\phi^i$ in the subset and can thus be denoted by $W(\Omega)$, 
and we define the fermionic subset weight
\begin{align}
\Sdet_{\mu,m}(\Omega) = \sum_{i=1}^{\Ns} {\det}^{N_f} \D{\mu}{m}(\phi^i) , 
\label{sigma}
\end{align}
which depends on $m$ and $\mu$.
The construction of the subsets will be described below, but its crucial property is that the sum $\Sdet_{\mu,m}(\Omega)$ over all the determinants in a subset is \textit{real} and \textit{positive}, for every $\Omega$ of the ensemble, and can thus be used as a probabilistic weight in a Monte Carlo method. 
The sample average of an observable $\Y$ over a sample of $N_\text{MC}$ subsets $\Omega_k$ is computed as
\begin{align}
  \overline \Y_{\mu,m} &= \frac{1}{N_\text{MC}} \sum_{k=1}^{N_\text{MC}}
\sum_{i=1}^{\Ns} \frac{{\det}^{N_f} \D{\mu}{m}(\phi^{ki})}{\Sdet_{\mu,m}(\Omega_k)} \, \Y_{\mu,m}(\phi^{ki}) ,
\label{avgY}
\end{align}
where $\phi^{ki}=(\phi^{ki}_1,\phi^{ki}_2) \in\Omega_k$. Methods based on partial integrations/summations have also been considered by other authors, see, e.g., refs.~\cite{Gocksch:1988iz,Kieu:1993gw,Anagnostopoulos:2001yb,Ambjorn:2002pz,Fodor:2007vv}.

We now sketch how the subsets are constructed. 
As the partition function \eqref{partfun} is real and positive (for $\mu<1$) the basic idea is to create subsets of matrices for which $\Sdet_{\mu,m}$ has the same property.
For $\mu=0$ the Dirac matrices are anti-Hermitian and their determinants are real and positive. When $\mu$ is increased the matrices become non-Hermitian, the determinants complex and the sign problem sets in. For $\mu=1$ and $m=0$ the ensemble becomes maximally non-Hermitian, the average phase factor is zero and the sign problem maximal.
For this limiting case we ideally want to find a partner configuration $\tphi=(\tphi_1,\tphi_2)$ for any given $\phi$ of the ensemble, such that their fermionic determinants explicitly cancel. This is easily achieved if,  given some angle $\theta$, we can find a $\psi$ for which $D_{1,0}(\psi)=e^{i\theta} D_{1,0}(\phi)$. Using \eqref{Ddef} the latter can be written as 
\begin{align}
e^{i\theta} D_{1,0}(\phi)
&=  \begin{pmatrix}
    0 & i\tphi_1 + \tphi_2 \\
    i \tphi_1^\dagger + \tphi_2^\dagger & 0
  \end{pmatrix}
 = \D{1}{0}(\tphi) ,
  \label{Dphimu=1}
\end{align}
where
\begin{align}
\begin{cases}
\tphi_1(\phi;\theta) = \cos\theta\;\phi_1 + \sin\theta\;\phi_2 \\
\tphi_2(\phi;\theta) = \cos\theta\;\phi_2 - \sin\theta\;\phi_1
\end{cases} 
\label{tphi}
\end{align}
are orthogonal rotations of $(\phi_1,\phi_2)$, which emerge when $e^{i\theta} D_{1,0}(\phi)$ is explicitly rewritten in an anti-Hermitian and a Hermitian part. From \eqref{rmtdis} it follows that $w(\tphi_1)w(\tphi_2)=w(\phi_1)w(\phi_2)$ such that $\tphi$ and $\phi$ have the same Gaussian weights in the partition function, for any rotation $\theta$. 
An exact pairwise cancellation of the fermionic determinants will happen for any $\theta$ obeying $e^{i2NN_f\theta}=-1$. 
The idea of canceling determinants can be extended from pairs $(\phi,\tphi)$ to subsets $\{\tphi(\phi;\theta_n)\}$, where the $\theta_n$ are such that the total sum of fermionic determinants cancels for $\mu=1$ and $m=0$, i.e., $\sum_n e^{i2NN_f\theta_n} = 0$.

This idea is now ported to arbitrary chemical potential $\mu$ and mass $m$.
In the following we will consider subsets
\begin{align}
\Omega(\phi)=\left\{ \tphi(\phi;\theta_n):  \theta_n = \frac{\pi n}{\Ns} \wedge n=0,\dotsc,\Ns\!-\!1 \right\} ,
\label{subset}
\end{align}
with $\tphi=(\tphi_1, \tphi_2)$ defined in \eqref{tphi} and $\Ns$ a positive integer. 
To each configuration $\phi=(\phi_1,\phi_2)$ of the random matrix ensemble corresponds a subset $\Omega(\phi)$, and the set of all subsets forms an $\Ns$-fold covering of the random matrix ensemble. 
The configurations $\tphi(\phi;\theta)$ have Gaussian weights independent of $\theta$, but different fermionic weights $\det^{N_f}\!\!\D{\mu}{m}(\tphi(\phi;\theta))$.

The method presented in this paper is based on the following \textit{theorem:} For any $\Omega$ constructed according to \eqref{subset}, and for arbitrary $\mu<1$ and $m$, the fermionic subset weight $\Sdet_{\mu,m}(\Omega)$ defined in \eqref{sigma}, i.e., the sum of the fermionic determinants of the $\Ns$ configurations $\tphi(\phi;\theta_n)$, is \textit{real} and \textit{positive} if $\Ns > NN_f$.

More specifically, for $m=0$ it can be shown that 
\begin{align}
\Sdet_{\mu,0}(\Omega) = (1-\mu^2)^{NN_f}\Sdet_{0,0}(\Omega),
\label{conject}
\end{align}
for any $\mu$. As $\Sdet_{0,0}$ is real and positive, $\Sdet_{\mu,0}$ is also real and positive for $\mu<1$. 
For $\mu=1$ the sum of determinants is exactly zero. For nonzero mass we can show that $\Sdet_{\mu,m}$ is real and $\Sdet_{\mu,m}(\Omega) > (1-\mu^2)^{NN_f}\Sdet_{0,m}(\Omega)$ for $\mu<1$, such that the weight $\Sdet_{\mu,m}$ is positive. 
The weights $\Sdet_{\mu,m}$ can thus be used to generate subsets of random matrices using a Metropolis algorithm and compute observables using \eqref{avgY}. In practice, $\Ns$ will be set to its minimum value, i.e., $\Ns=N N_f+1$. 

This theorem can be proven analytically and was thoroughly tested numerically. The proof will be given in a forthcoming paper.

We applied the subset method to compute the chiral condensate of the chiral random matrix ensemble, which is defined as $\Sigma=\frac{1}{2N} \tr D^{-1}$ \cite{Osborn:2008jp}. 
We use the Metropolis algorithm to generate subsets $\Omega$ according to their weights $W(\Omega)\Sdet_{\mu,m}(\Omega)$, where the determinants are computed numerically. 
Successive subsets in the Markov chain are generated as follows: randomly choose a configuration in the current subset, generate a new configuration by making a random step, construct the subset corresponding to this new configuration, apply an accept-reject step to the newly proposed subset. In each Markov chain we generate 100,000 subsets. The chiral condensate is computed for each configuration and its average is computed using \eqref{avgY}. As usual, successive configurations in the Markov chains are correlated and the number of independent subsets is smaller by a factor $2\tau$, where $\tau$ is the integrated autocorrelation time. The statistical errors on the measurements are determined using the standard error formula corrected for the autocorrelations. 

To compare the subset method with standard reweighting methods, all the simulations 
were repeated using quenched, phase quenched and sign quenched reweighting \cite{Bloch:2011jk}. In those cases, we used $(N N_f+1)\times 100,000$ random matrices in the Markov chains, such that the total number of matrices generated in the reweighting methods is the same as in the subset method. For the sake of clarity, we only show the results of phase quenched reweighting in the figures below, as its results are representative for the various reweighting methods.

\begin{figure*}
\includegraphics{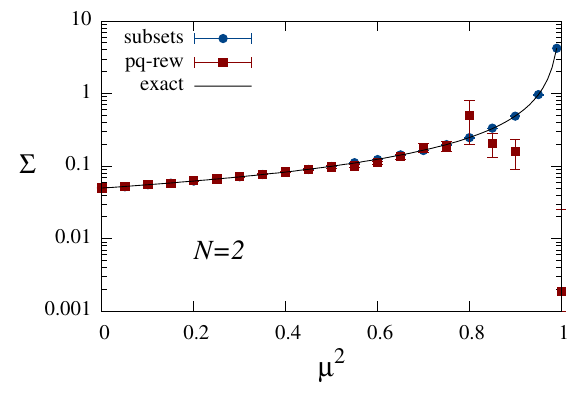}%
\includegraphics{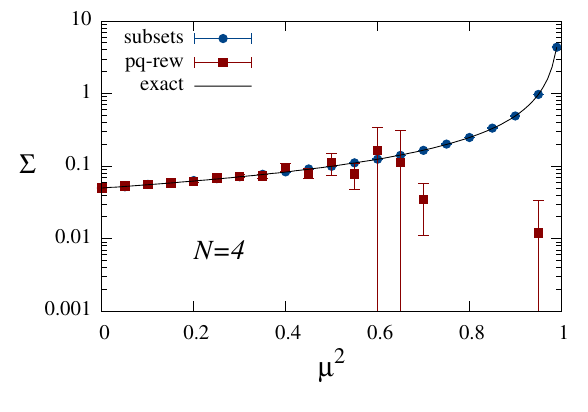}%
\includegraphics{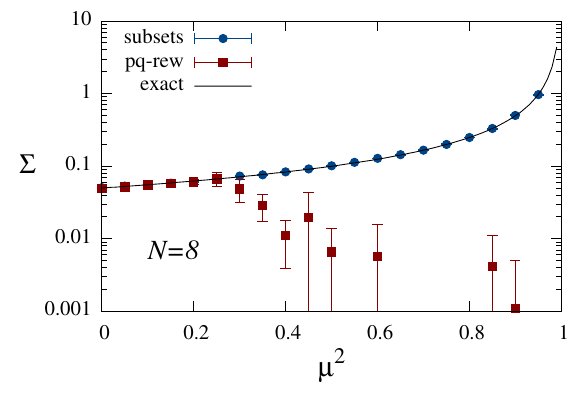}\\[-1mm]
\includegraphics{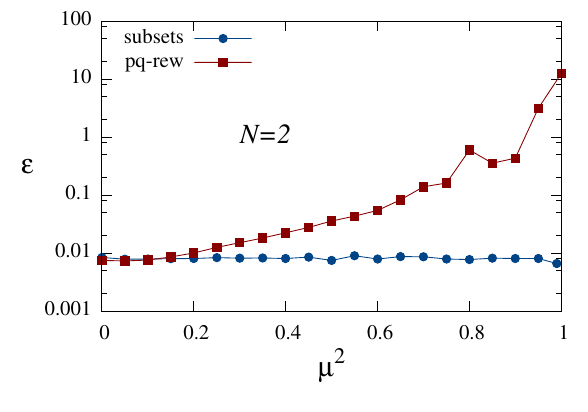}%
\includegraphics{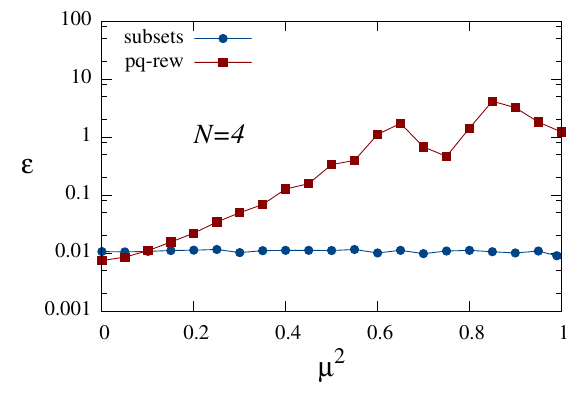}%
\includegraphics{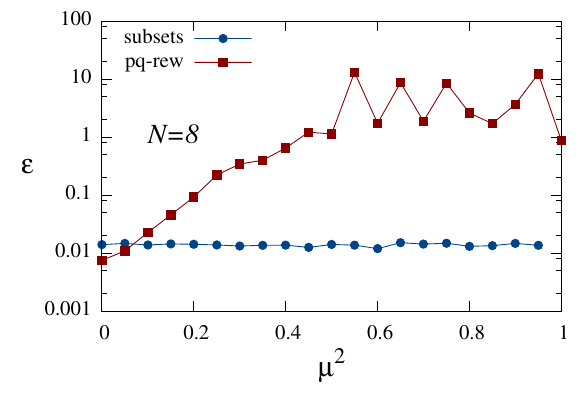}\vspace{-2mm}
\caption{Chiral condensate $\Sigma$ (top row) as a function of the chemical potential $\mu^2$ for the subset method and the phase-quenched reweighting method for $N=2,4,8$. The full line shows the exact analytical result of ref.~\cite{Osborn:2008jp}. The reweighting method fails for ever smaller $\mu^2$ when $N$ grows (some data points are negative and are left out of the semi-log plots). The corresponding relative statistical error $\varepsilon$ is shown in the bottom row. 
For the subset method, the error is independent of the chemical potential. 
The error for the reweighting method grows very rapidly and should only be trusted as long as the method works (see top row).\vspace{-2mm}}
\label{fig-cc-vs-mu}
\end{figure*}

We performed simulations for $N=2,\dotsc,34$, choosing $N_f=1$ and $m=0.1/2N$, so that the mass is small w.r.t. the magnitude of the smallest eigenvalue.
In fig.~\ref{fig-cc-vs-mu} the chiral condensate is shown as a function of the chemical potential for matrices with $N=2,4,8$. We compare the results obtained using the subset method with those from phase quenched reweighting. We also show the exact results computed in ref.~\cite{Osborn:2008jp}. 
The data are shown in the top row and the corresponding relative statistical errors in the bottom row. 
As the matrix size increases the reweighting method fails for smaller and smaller $\mu^2$ due to the sign and the overlap problem. As expected, the error of the reweighting method grows exponentially, until the method fails when the set of sampled matrices no longer overlaps with the relevant configurations for the given value of $\mu^2$ and $N$. 
This strongly contrasts with the results of the subset method which are reliable up to much larger values of $\mu^2$ and agree with the analytical predictions. 
Interestingly, the accuracy of the measurements is independent of the chemical potential. 

\begin{figure*}
\includegraphics{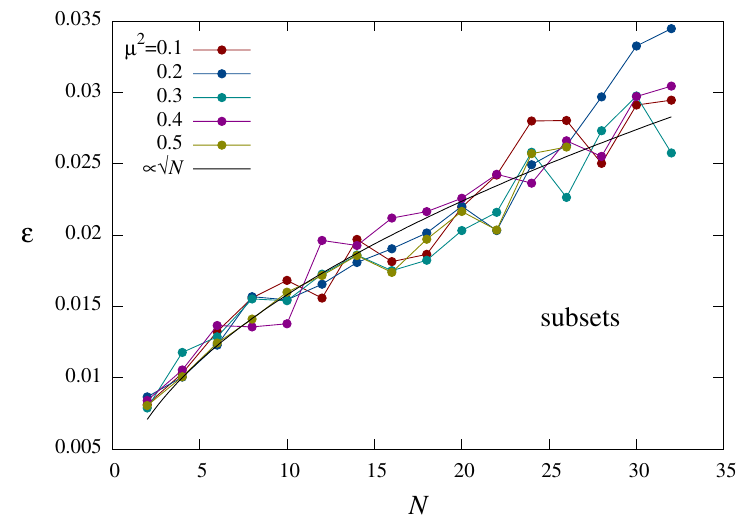}
\includegraphics{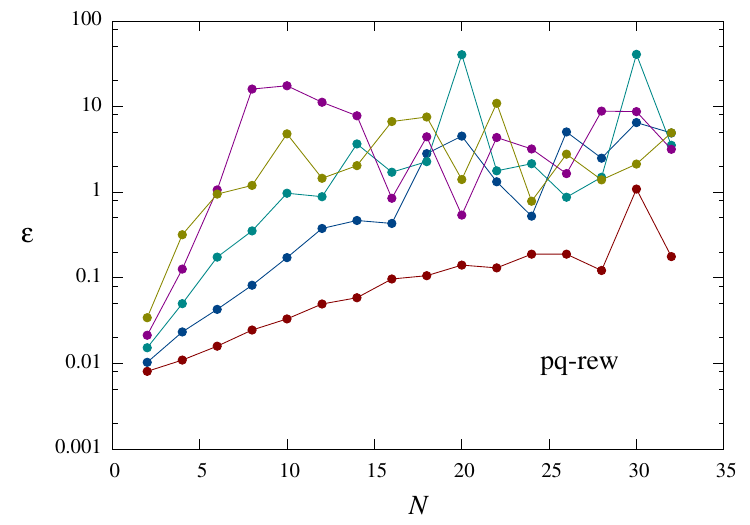}
\caption{Relative error $\varepsilon$ on the chiral condensate versus matrix size $N$ for various values of chemical potential $\mu^2=0.1,0.2,0.3,0.4,0.5$. The left plot shows the results of the subset method. The full curve $\varepsilon(N) \propto \sqrt{N}$ serves to guide the eye. As a comparison, the right plot shows the relative error for phase quenched reweighting on a semi-log plot (the color coding for $\mu^2$ is the same as in the left plot). The error increases exponentially with $N$, until the reweighting method fails.}
\label{fig-cc-vs-N}
\end{figure*}

We also measured the chiral condensate as a function of $N$ for fixed values of $\mu^2$. The $N$-dependence of the relative statistical error on the chiral condensate is shown in fig.~\ref{fig-cc-vs-N} (left: subset method, right: phase quenched reweighting) for different values of $\mu^2$. 
For a fixed number of subsets the error in the subset method (left panel) increases approximately as $\sqrt{N}$ and is independent of $\mu$ (the latter also follows from fig.~\ref{fig-cc-vs-mu}). 
If we fix the number of matrices, rather than the number of subsets, the error will increase with an additional factor $\sqrt{N}$ (as the subset size itself grows with $N+1$), such that the overall relative error will grow linearly with $N$.
Turning the argument around, to achieve a constant error the number of subsets would have to grow proportionally to $N$, i.e., the total number of matrices should approximately grow as $N^2$. The right plot shows the same quantity for phase quenched reweighting (on semi-log scale). We observe that the error grows exponentially with $N$ until the method completely fails when the error stagnates around one and is no longer reliable. Note that for both methods the additional cost for the numerical computation of the determinants is proportional to $N^3$.

\begin{figure}
\includegraphics{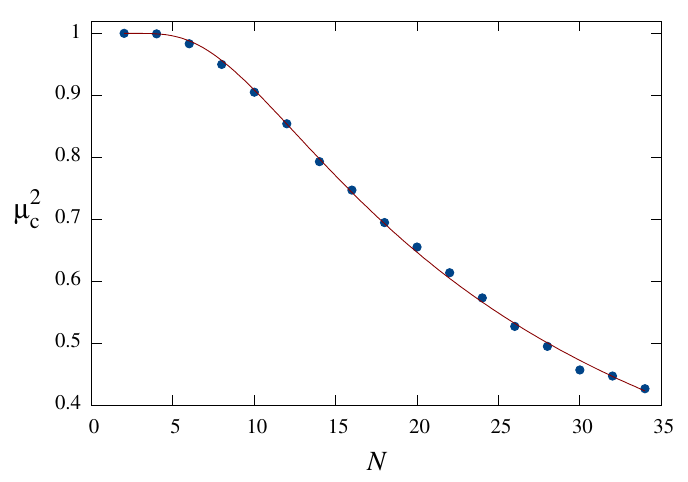}
\caption{Chemical potential $\mu^2_{c}(N)$ where the subset method breaks down due to the finite precision in the computation of $\Sdet_{\mu,m}$.
 The full line shows the fit $\mu_c^2 = 1-\exp(-38/N^{1.2})$. }
\label{fig-muc}
\end{figure}

For large values of $N$ or $\mu^2$ we observed that the subset method breaks down at an $N$-dependent value $\mu_c(N)$, because the finite precision arithmetic limits the accuracy of $\Sdet_{\mu,m}$, which can lead to a violation of its positivity. The breakdown occurs when the sum of the determinants in a subset becomes of the order of the accuracy of the individual determinants. The approximate value $\mu^2_c$ is shown as a function of $N$ in fig.~\ref{fig-muc} for simulations performed in double precision arithmetics. The data are fitted well by $\mu_c^2 \approx 1-\exp(-38/N^{1.2})$, whose functional form is inspired by \eqref{conject}. For large $N$ the fitted curve goes like $1/N^{1.2}$.
The breakdown can easily be detected during simulations by monitoring $\Sdet_{\mu,m}$ and the magnitude of the individual determinants. When the ratio of both quantities is of the order of the machine precision the method no longer gives sensible results. Usually this is accompanied by a numerical violation of the positivity of $\Sdet_{\mu,m}$.
For simulations at $m=0$ this breakdown can be avoided altogether by using the analytic formula \eqref{conject} to compute the sampling weight $\Sdet_{\mu,0}$ from  $\Sdet_{0,0}$.

Note that the breakdown caused by the finite numerical accuracy will equally well show up in standard reweighting methods, even if an exponentially large effort is invested to keep the statistical error under control.
This accuracy problem is therefore different from the sign problem, as the latter is of a pure statistical nature and does not depend on the use of finite-precision or exact arithmetic. The former is an additional problem of numerical nature which can be improved upon if necessary by using more sophisticated numerical methods. 

What happened to the sign problem in the subset method? In standard reweighting methods the large cancellations, inherent to simulations at real chemical potential, happen through statistical sampling of the partition function.
In the subset method these cancellations are removed from the statistical sampling procedure and become confined inside the subsets, which are constructed in a deterministic way and whose partial sums $\Sdet_{\mu,m}$ yield net real and positive weights. 
Thus, a fundamental difference between both methods is that the number of configurations in reweighting grows exponentially with the volume to maintain the necessary statistical accuracy on the average weight factor \cite{deForcrand:2010ys}, whereas only $N N_f+1$ matrices per subset are needed in the subset method to compute the positive subset weights $\Sdet_{\mu,m}$, independently of its magnitude.
This is how the subset method avoids the large statistical cancellations characterizing the sign problem.

From a practical point of view, even though finite-precision arithmetic in the numerical simulations eventually leads to a breakdown of the subset method, our numerical tests have shown that the method yields a vast improvement over the standard reweighting methods for the dynamical simulation of random matrices. 
With the new method higher values of $\mu^2$ and $N$ can be accessed, without any loss of accuracy when increasing the chemical potential and with a measurement error growing proportionally to $N$ when increasing the matrix size.

The crucial question whether the subset method is also applicable to physically relevant systems will be investigated in future research.

I would like to thank Philippe de Forcrand and Tilo Wettig for useful discussions. This work was supported by the DFG collaborative research center SFB/TR--55.

\sloppypar
\bibliography{biblio} 

\end{document}